\documentclass[aps,nofootinbib,onecolumn,showpacs,10pt]{revtex4}
\usepackage{graphicx,epsfig,amssymb}
\usepackage{epsf}
\usepackage{graphicx}

\def   \lsth     {littlest Higgs}

\newcommand\iden{\leavevmode\hbox{\small1\normalsize\kern-.33em1}}
\def \nn {\nonumber}
\def\w{{\rm w}}

\def\gmm{{\gamma_{\mu}}}

\def\gff{{\gamma_{5}}}

\def\gvi{{g_{V_i}}}
\def\gai{{g_{A_i}}}

\def\dwi{{M_i \Gamma_i}}

\begin{document}
\vspace*{2cm}
\title{Pair production of charged scalars and lepton flavor violating signals in the littlest Higgs model  at $e^{+}e^{-}$ colliders }
\author{A. \c{C}a\~{g}{\i}l\footnote{e-mail: ayse.cagil@cern.ch}}
\affiliation{\vspace*{0.1in} Physics Department, Middle East
Technical University,\\06531 Ankara, Turkey}
%
%%%%%%%%%%%%%%%%%%%%%%%%%%%%%%%%%%%%%%%%%%%%%%%%%%%%%%%%%%%%%%
% You may repeat \author \address as often as necessary      %
%%%%%%%%%%%%%%%%%%%%%%%%%%%%%%%%%%%%%%%%%%%%%%%%%%%%%%%%%%%%%%
%\vspace*{-1.7cm}
\vspace*{1.0cm}
%
%
%
%\setlength{\baselineskip}{24pt} \setlength{\baselineskip}{7mm}
%%%%%%%%%%%%%%%%%%%%%%%%%%%%%%%%%%%%%%%%%%%%%%%%%%%%%%%%%%%%%%%%%%%%%%%%%%%%%%%%%%%%%%%%
\begin{abstract}
In this work pair productions of charged and doubly charged scalars in
the framework of littlest Higgs model at $e^+e^-$ colliders are studied. In the allowed parameter space of the littlest Higgs model, the production rates of the scalar pairs are calculated. It is obtained that pair productions of charged and doubly charged scalars are reachable at $e^+ e^-$ colliders with energy $\sqrt{S}\geq 1.7TeV$. Using the lepton flavor violating decays of charged scalars calculated in literature, final state analysis is done for pair production processes. This analysis show that depending on the model parameters, lepton number and lepton flavor violations can be observed free from any backgrounds.
\end{abstract}
 \maketitle
%\pagenumbering{arabic} \setcounter{page}{1}
%%%
%%%%%%%%%%%%%%%%%%%%%%%%%%%%%%%%%%%%%%%%%%%%%%%%%%%%%%%%%%%%%%%%%%%%%%%
\section{Introduction}
%%%%%%%%%%%%%%%%%%%%%%%%%%%%%%%%%%%%%%%%%%%%%%%%%%%%%%%%%%%%%%%%%%%%%%%
Despite the impressive success of the Standard Model (SM) in
describing all experimental data, it contains many unsolved
problems, such as, origin of mass and CP violation, hierarchy
problem, number of generations, baryon anti-baryon asymmetry, etc.
For solution of these problems it is necessary to go beyond SM. To obtain a ``natural''  solution 
to hierarchy problem little Higgs models \cite{lh1,lhmodels1,lhmodels2,lhmodels3} are introduced. 
In little Higgs models new physics is introduced at $TeV$ scale; these models predict existence of new 
particles and new interactions of these particles with SM particles as well as interactions among themselves. 
The phenomenologies of the little Higgs models are reviewed widely\cite{perelstein1,thanrev,schmalzrev1}, and the 
constraints on little Higgs models are studied
\cite{csaki1,perelstein2ew,B1rizzo,Bdawson,Bkilian,Bdias,B2csaki}. 

Among little Higgs models, there are several variations which differ in the assumed higher symmetry group 
and in the representation of the scalar multiplets, one of which is the littlest Higgs model\cite{lh1}. 
The existence of new particles in littlest Higgs model includes new heavy gauge bosons and a new scalar sector consisting of 
a neutral heavy scalar ($\phi^0$), a neutral pseudo-scalar ($\phi^P$), a singly charged scalar ($\phi^+$) and a doubly charged 
scalar ($\phi^{++}$) receives special attention. The charged scalars $\phi^+$ and  $\phi^{++}$  of the littlest Higgs 
model are under special interest since they have distinct signatures in future colliders. Productions of 
charged scalars and their signatures at ILC, LHC and THERA are studied via $e^-\gamma \to e^+ \phi^{--}$, $e^- p\to l^+ X \phi^{--}$ 
and $\bar{q}q' \to \phi^+ \phi^{--}$ processes\cite{ch1ref}.   Also $Z_L$ associated productions and final 
collider signatures of charged scalars are studied via $e^+e^-\to Z_L \phi^+ \phi^-$ and $e^+e^-\to Z_L \phi^{++} \phi^{--}$ processes at a linear collider\cite{cagil1}.

In this work, the productions of single and doubly charged
scalar pairs  via $e^+e^-\to \phi^+\phi^-$ and $e^+e^-\to \phi^{++}\phi^{--}$
 processes at future $e^{-}e^{+}$ colliders, namely,
International Linear Collider (ILC) \cite{ILC} and Compact Linear
Collider (CLIC) \cite{CLIC} are examined. The dependence of cross
sections to the littlest Higgs model parameters at the range allowed
by electroweak precision observables are calculated. It is found that the production
rates of the single charged scalar pairs  are less than the
production rates of doubly charged scalar pairs, but both channels
will be achieved at future $e^{-}e^{+}$ colliders at $\sqrt{s}\geq
1.7GeV$. In addition to production rates, the final signatures of the productions are also analyzed considering the lepton flavor violating decays of the charged and the double charged scalars, whose branching ratios have been studied by T.Han et al\cite{thanlept1}. It is found that pair productions of charged and double charged scalars lead to distinct signatures in $e^+e^-$ colliders including lepton number and lepton flavor violating ones.

The paper is organized as follows: In section $2$, the cross sections of
pair productions of charged scalars at $e^{-}e^{+}$ colliders are calculated. Section $3$ contains our numerical results and discussions.
%
%%%%%%%%%%%%%%%%%%%%%%%%%%%%%%%%%%%%%%%%%%%%%%%%%%%%%%%%%%%%%%%%%%%%%%%%%
\section{Theoretical Framework}
%%%%%%%%%%%%%%%%%%%%%%%%%%%%%%%%%%%%%%%%%%%%%%%%%%%%%%%%%%%%%%%%%%%%%%%%%%
Before examining the pair productions of the charged scalars, we remind the main ingredients of the littlest Higgs model and lepton 
flavor violation in littlest Higgs model. The littlest Higgs model assumes a higher symmetry group $SU(5)$ 
with a weakly gauged subgroup of $(SU(2)\otimes U(1))^2$. Among consecutive symmetry breakings 
first $SU(5)$ is broken to $SO(5)$ at $TeV$ scale, and simultaneously subgauged group $(SU(2)\otimes U(1))^2$ is broken to 
$SU(2)\otimes U(1)$. Then at $v\sim246GeV$ ordinary electroweak symmetry breaking (EWSB) occurs. As a result of 
higher symmetry breaking new scalar sector enters the model, which is the scalar triplet at the end whose members are $\phi^0$, $\phi^P$, $\phi^+$ and $\phi^{++}$. 
Also from symmetry breaking of gauged group $(SU(2)\otimes U(1))^2$ new bosons $A_H$, $Z_H$ and $W_H$ gain mass. 

In summary, littlest Higgs model contain four physical scalars; Higgs scalar: $H$, new heavy scalars: 
$\phi^0$, $\phi^+$ and $\phi^{++}$, and a new heavy pseudo-scalar: $\phi^P$. All scalars excluding $H$ are degenerate in mass:

\begin{eqnarray}
    M_\phi =\frac{\sqrt{2} f}{v\sqrt{1-(\frac{4 v'
    f}{v^2})^2}}M_H,
\end{eqnarray}
where $M_H$ is the mass of the Higgs boson, $f$ is the higher symmetry breaking scale of the littlest Higgs model, $v$ and $v'$ are the vacuum expectation values (VEVs) of the Higgs field 
and the scalar triplet respectively. The vacuum expectation values of Higgs field and
scalar triplet are given as;
\begin{equation}\label{vandvprimelimits}
\langle h^0 \rangle = v/\sqrt{2}~~,~~~~\langle i \phi^0 \rangle =
v^{\prime}\leq \frac{v^2}{4f},
\end{equation}
bounded by electroweak precision data, where $v=246 GeV$. 

In the littlest Higgs model, the masses of the new gauge bosons $A_H$, $Z_H$ and $W_H$ are given as\cite{thanrev}:
\begin{eqnarray}\label{massesvectors}
    M_{A_L}^2 &=& 0 ,\nonumber \\
    M_{Z_L}^2 &=& m_z^2
    \left[ 1 - \frac{v^2}{f^2} \left( \frac{1}{6}
    + \frac{1}{4} (c^2-s^2)^2
    + \frac{5}{4} (c^{\prime 2}-s^{\prime 2})^2 \right)
    + 8 \frac{v^{\prime 2}}{v^2} \right],
    \nonumber \\
   \nn M_{A_H}^2 &=&
    \frac{f^2 g^{\prime 2}}{20 s^{\prime 2} c^{\prime 2}}
    - \frac{1}{4} g^{\prime 2} v^2 + g^2 v^2 \frac{x_H}{4s^2c^2}\\
\nn& =& m_z^2 s_{\w}^2 \left(
    \frac{ f^2 }{5 s^{\prime 2} c^{\prime 2}v^2}
    - 1 + \frac{x_H c_{\w}^2}{4s^2c^2  s_{\w}^2} \right),
    \nonumber \\
    M_{Z_H}^2 &=& \frac{f^2g^2}{4s^2c^2}
    - \frac{1}{4} g^2 v^2
    - g^{\prime 2} v^2 \frac{x_H}{4s^{\prime 2}c^{\prime 2}}\\
   \nn & =& m_w^2 \left( \frac{f^2}{s^2c^2 v^2}
    - 1 -  \frac{x_H s_{\w}^2}{s^{\prime 2}c^{\prime 2}c_{\w}^2}\right) ,
\end{eqnarray}
where   $m_z\equiv {gv}/(2c_{\w})$ and $x_H = \frac{5}{2} g
g^{\prime}
    \frac{scs^{\prime}c^{\prime} (c^2s^{\prime 2} + s^2c^{\prime 2})}
    {(5g^2 s^{\prime 2} c^{\prime 2} - g^{\prime 2} s^2 c^2)}$,
$s_{\w}$ and $c_{\w}$ are the usual SM weak mixing angles,
$c_{\w}\equiv\cos(\theta_{\w})$ and $c_{\w}^\prime\equiv\cos(\theta_{\w}^\prime)$. 
In Eq. \ref{massesvectors}, $s\equiv \sin(\theta)$ and $s'\equiv \sin(\theta')$ are the mixing angles of $SU(2)$ and $U(1)$ subgroups respectively.

One of the interesting features of littlest Higgs 
model is that as a result of the extended Higgs sector, it predicts lepton flavor violation by unit two by implementing a Majorana type mass in Yukawa
lagrangian\cite{thanlept1,gaurlept1,cinlept_L2yue,gaurlept2}, such as:
\begin{equation}\label{lepviol1}
    {\cal L}_{LFV} = iY_{ij} L_i^T \ \phi \, C^{-1} L_j + {\rm h.c.},
\end{equation}
where $L_i$ are the lepton doublets $\left(
                                       \begin{array}{cc}
                                         l &\nu_l \\
                                       \end{array}
                                     \right)$,
and $Y_{ij}$ is the Yukawa coupling with $Y_{ii}=Y$ and $Y_{ij(i\neq
j)}=Y'$ . The values of Yukawa couplings $Y$ and $Y'$ are restricted
by the current constraints on the neutrino
masses\cite{neutrinomass}, given as; $M_{ij}=Y_{ij}v'\simeq
10^{-10}GeV$. Since the vacuum expectation value $v'$ has only an
upper bound; $v'<\frac{v^2}{4f}$, $Y_{ij}$ can be taken up to order of unity
without making $v'$ unnaturally small. In this work the values of
the Yukawa mixings are taken to be $10^{-10}\leq Y\leq 1$, 
$Y'\sim 10^{-10}$, and the vacuum expectation value $1GeV\geq v'\geq10^{-10}
GeV$.

%%%%%%%%%%%%5

In littlest Higgs model, the symmetry braking scale $f$ and mixing angles $s$ and $s'$ are free parameters and they are constrained by observables\cite{csaki1,perelstein2ew,B1rizzo,Bdawson,Bkilian,Bdias,B2csaki}. 
The recent data from Tevatron and LEPII constrain the mass of the lightest heavy vector boson as $M_{A_H}\gtrsim900GeV$\cite{Bdawson,TEVATRON}. In the original formulation of the littlest Higgs model, these data imposes strong constraints on symmetry breaking scale($f>3.5 - 4 TeV$). But in this work by gauging fermions in both $U(1)$ subgroups, fermion boson couplings are modified as done in\cite{B2csaki}. 
With this modification the symmetry breaking scale can be lowered to $f=1TeV$, which allows the mass of the $A_H$ to be at the order of few $GeV$s. In this case the allowed parameter region of the littlest Higgs 
model is  as follows. For low values of the symmetry breaking scale $1TeV
\leq f \leq 2 TeV$, mixing angles $s$ and $s'$ between gauge bosons
are in the range $0.8\leq s \leq 1$ and $0.6\leq s' \leq 0.7 $, and
for $ 2TeV \leq f \leq 3TeV$ they have acceptable values in the
range $0.65\leq s \leq 1$ and $0.4\leq s' \leq 0.9$. For the higher
values of the symmetry breaking scale $f\geq 3TeV$, the mixing
angles are less constrained, since the corrections to SM observables
from {\lsth} model comes in the form $\frac{v^2}{f^2}$ and higher
orders.

%%%%%%%%%5
For calculation of production rates  and decays of charged scalar pairs,
the couplings between fermions and neutral vector bosons are needed.
The couplings of fermions with gauge bosons are written as $i \gmm
(\gvi + \gai \gff)$ where $i=1,2,3,4$ corresponds to $Z_L$, $Z_H$,
$A_H$ and $A_L$ respectively. These couplings of vectors with
$e^{-}e^{+}$ are given in table \ref{gVgA}, where $e=\sqrt{4 \pi\alpha}$,
$y_e=\frac{3}{5}$ for anomaly cancelations, $x_Z^{W^{\prime}} =
-\frac{1}{2c_{\w}} sc(c^2-s^2)$ and $x_Z^{B^{\prime}} =
-\frac{5}{2s_{\w}} s^{\prime}c^{\prime}
    (c^{\prime 2}-s^{\prime 2})$. It is seen from table
\ref{gVgA} that vector and axial vector couplings of SM $Z_L
e^{-}e^{+}$ vertex also gets contributions from littlest Higgs
model. As a result total decay widths of SM vector bosons also gets
corrections of the order $\frac{v^2}{f^2}$, since the decay widths
of vectors to fermion couples are written as; $\Gamma(V_i\rightarrow
f\bar{f})=\frac{N}{24\pi}(g^2_V+g^2_A)M_{V_i}$ where $N=3$ for
quarks, and $N=1$ for fermions. Total decay widths of new vectors
$A_H$ and $Z_H$ are given as \cite{A3}:
\begin{equation}
\Gamma_{A_H}\approx  \frac{g'^2 M_{A_H}(21-70 s'^2 +59 s'^4)}{48 \pi
s'^2 (1-s'^2) }~~,~~~~ \nn\Gamma_{Z_{H}}\approx \frac{g^2 (193 - 388
s^2 + 196 s^4)}{768 \pi s^2 (s^2 -1)}M_{Z_H}.
\end{equation}
\begin{table}[hbt]
\begin{center}
\begin{tabular}{|c||c|c|c|}
  \hline
  % after \\: \hline or \cline{col1-col2} \cline{col3-col4} ...
   $i$& vertexes &$\gvi$  &$\gai$    \\
  \hline\hline 1&$e\bar{e} Z_L$ &
    $-\frac{g}{2c_{\w}} \left\{ (-\frac{1}{2} + 2 s^2_{\w})- \frac{v^2}{f^2} \left[ -c_{\w} x_Z^{W^{\prime}} c/2s
    \right. \right.$ &
    $-\frac{g}{2c_{\w}} \left\{ \frac{1}{2}
    - \frac{v^2}{f^2} \left[ c_{\w} x_Z^{W^{\prime}} c/2s
    \right. \right.$ \\
& &  $\left. \left.
    + \frac{s_{\w} x_Z^{B^{\prime}}}{s^{\prime}c^{\prime}}
    \left( 2y_e - \frac{9}{5} + \frac{3}{2} c^{\prime 2}
    \right) \right] \right\}$&
    $\left. \left.
    + \frac{s_{\w} x_Z^{B^{\prime}}}{s^{\prime}c^{\prime}}
    \left( -\frac{1}{5} + \frac{1}{2} c^{\prime 2} \right)
    \right] \right\}$ \\
  \hline 2& $e\bar{e} Z_H$ &$-gc/4s$ & $gc/4s$      \\
  \hline 3&$e\bar{e} A_H$  &
    $\frac{g^{\prime}}{2s^{\prime}c^{\prime}}
    \left( 2y_e - \frac{9}{5} + \frac{3}{2} c^{\prime 2} \right)$ &
    $\frac{g^{\prime}}{2s^{\prime}c^{\prime}}
    \left( -\frac{1}{5} + \frac{1}{2} c^{\prime 2} \right)$    \\
  \hline4& $e\bar{e} A_L$&e&0     \\
  \hline

\end{tabular}
\caption{The vector and axial vector couplings of $e\bar{e}$ with
vector bosons. Feynman rules for $e\bar{e} V_i$ vertexes are given
as $i \gmm (\gvi + \gai \gff)$ \cite{thanrev}.}\label{gVgA}
\end{center}

\end{table}

The new scalars and pseudoscalar also contribute to the analysis as
taken into account in this study. The decay modes of littlest Higgs model scalars, including the lepton flavor violating decays, are studied in T.Han et al\cite{thanlept1}. Since these new scalars have lepton flavor
violating modes, their total widths will depend on the Yukawa
couplings $Y_{ii}=Y$ and $Y_{ij(i\neq j)}=Y'$. The decay width of
$\phi^{++}$ is given as\cite{thanlept1}:
\begin{eqnarray}\label{dwp2}
 \nn   \Gamma_{\phi^{++}}&=&\Gamma (W^+_L W^+_L)+3 \Gamma ( \ell^+_i \ell^+_i) +3 \Gamma (\ell^+_i \ell^+_j )\\
    &\approx&  \frac{v^{\prime 2} M_{\phi}^3}{2 \pi v^4}+\frac{3}{8\pi } |Y|^2 M_\phi+\frac{3}{4\pi } |Y'|^2 M_\phi
\end{eqnarray}
For the single charged scalar, the decay width is given
by\cite{thanlept1}:
\begin{eqnarray}\label{dwp1}
\nn \Gamma_{\phi^+}&=&3 \Gamma ( \ell^+_i \bar\nu_i)+6 \Gamma
(\ell^+_i \bar\nu_j)+\Gamma ( W_L^+ H)+\Gamma
(W_L^+ Z_L)+\Gamma (t \bar{b})+\Gamma (T \bar{b})\\
 &\approx & \frac{N_c M^2_t M_\phi }{32 \pi f^2}+\frac{v^{\prime
2}M^3_\phi }{2\pi v^4} +\frac{3}{8\pi
    }
    |Y|^2 M_\phi+\frac{3}{4\pi
    }
    |Y'|^2 M_\phi.
\end{eqnarray}
%%%%%%%%%%%%%%%%%%%%%%%
In this final expression, the decay of single charged scalar to $T\bar{b}$ is neglected since in the parameter space considered in this work, $M_\phi \sim M_T$, hence this decay is suppressed. It is seen from the decay widths of the scalars that lepton number
violation is proportional to  $|Y|^2$ if the final state leptons are
from the same family and to $|Y'|^2$ for final state leptons are
from different generations . The branching ratios of scalars
decaying to same family of leptons is denoted as $BR[Y]$ and to
leptons of different flavor is $BR[Y']$.
\begin{figure}[tbh]
\begin{center}
\includegraphics[width=5cm]{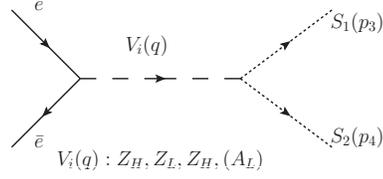}
\qquad\allowbreak \vskip2mm
\end{center}\caption{Feynman diagrams
contributing to $e^{-}e^{+}\rightarrow S_1 S_2$ processes in {\lsth}
model. $S_1\&S_2:  \phi^+ \& \phi^- ,\phi^{++} \& \phi^{--}$.}
\label{fdSS}
\end{figure}

The scalar pairs produced at $e^+e^-$ colliders is described by
Feynman diagrams presented in Fig. \ref{fdSS}.The necessary vertices
for considered processes are given in tables \ref{PP11Vcouplings}
and \ref{PP22Vcouplings} for single charged pair production and
doubly charged pair production of scalars respectively.
\begin{table}[htb]
\begin{center}
\begin{tabular}{|c||c|c|}
  \hline
  % after \\: \hline or \cline{col1-col2} \cline{col3-col4} ...
  i/j&vertices & i $E^{\phi\phi}_{i} Q_\mu$\\
\hline\hline  1 &$\phi^+ \phi^- Z_L$&$i \frac{g}{c_\w} s^2_\w (p_1-p_2)_{\mu}$   \\
\hline  2&$\phi^+ \phi^- Z_H$ &$\mathcal{O}(v^2/f^2)\sim 0$  \\
\hline  3&$\phi^+ \phi^- A_H$ & $i
    g^{\prime} \frac{(c^{\prime 2}-s^{\prime 2})}{2s^{\prime}c^{\prime}}
    (p_1-p_2)_{\mu}$ \\
\hline  4&$\phi^+ \phi^- A_L$ & $-i e (p_1 - p_2)_{\mu}$ \\
\hline
\end{tabular}

\caption{The interaction vertices for $\phi^+ \phi^- V_i $ vertices.
Their couplings are given in the form $i E^{\phi\phi}_{i} Q_{\mu}$,
where $Q_{\mu} =(p_1 -p_2)_\mu$ is the difference of momentums of
two charged scalars\cite{thanrev}.} \label{PP11Vcouplings}
\end{center}
\end{table}
\begin{table}[htb]
\begin{center}
\begin{tabular}{|c||c|c||c||c|c|}
  \hline
  % after \\: \hline or \cline{col1-col2} \cline{col3-col4} ...
  i/j&vertices & $i E'^{\phi\phi}_{i} Q_\mu$ \\
\hline\hline  1 &$\phi^{++} \phi^{--}Z_L$&$-i \frac{g}{c_\w} \left( 1 - 2 s^2_\w \right) (p_1-p_2)_{\mu}$   \\
\hline  2&$\phi^{++} \phi^{--}Z_H$ &$i g \frac{(c^2-s^2)}{2sc} (p_1-p_2)_{\mu}$  \\
\hline  3&$\phi^{++} \phi^{--} A_H$ &$i
    g^{\prime} \frac{(c^{\prime 2}-s^{\prime 2})}{2s^{\prime}c^{\prime}}
    (p_1-p_2)_{\mu}$  \\
\hline  4&$\phi^{++} \phi^{--} A_L$ &$-2 i e (p_1-p_2)_{\mu}$  \\
\hline
\end{tabular}

\caption{The interaction vertices for $\phi^{++} \phi^{--} V_i $
vertices. Their couplings are given in the form $i E'^{\phi\phi}_{i}
Q_{\mu}$, where $Q_{\mu} = (p_1-p_2)_\mu$ is the difference of
momentums of the scalars\cite{thanrev}.} \label{PP22Vcouplings}
\end{center}
\end{table}
The amplitudes corresponding to these Feynman diagrams are written
as:
\begin{equation}\label{Mc1i}
M_{1}=\bar{u}[-p_2] i \gmm  g_{V_4}u[p_1](i)
    \frac{g^{\alpha\mu}}{q^2}i E_{4} (p_3 - p_4)_{\alpha},
\end{equation}
\begin{equation}\label{Mc2i}
M_{2}=\sum_{i=1}^{3}\bar{u}[-p_2] i \gmm  (\gvi +\gai \gff)u[p_1](i)
\frac{g^{\alpha\mu}-\frac{q^{\mu}q^{\alpha}}{M^{2}_{i}}}{q^2-M^{2}_{i}+i\dwi}
i E_{i} (p_3 - p_4)_{\alpha},
\end{equation}
where $q=p_1+p_2$ and coefficients
$E_{i}=E_i^{\phi\phi}(E_i^{\prime\phi\phi})$ are given in table
\ref{PP11Vcouplings}(\ref{PP22Vcouplings}) for single(doubly)
charged scalar pair and $M_1$ is for the photon propagating diagram.

The pair production cross sections are calculated by squaring the amplitudes and integrating over phase space. The analytical expression for cross section is found as

\begin{eqnarray}\label{CSs}
 \nn\sigma (S)&=&\frac{\sqrt{S^2-4 S M_\phi^2}}{96 \pi}\left[\left(\frac{2 E_4 g_{V_4}}{S}\right)^2+\sum_{i=1}^{3}\frac{8 g_{V_4}g_{V_i}E_i E_4}{S(S-M_i^2)}\right.\\
&&\left. +\sum_{i,j=1}^{3}\frac{4 E_i E_j (g_{V_i}g_{V_j}+g_{A_i}g_{A_j})}{(S-M_i^2)(S-M_j^2)}\right],
\end{eqnarray}
where $S=q^2$ and  $E_{i}=E_i^{\phi\phi}(E_i^{\prime\phi\phi})$ for the process $e^+e^-\to\phi^{+}\phi^{-}$ ($e^+e^-\to\phi^{++}\phi^{--}$). 

%
%%%%%%%%%%%%%%%%%%%%%%%%%%%%5
%\newpage
%
\section{The results and discussion}\label{sectiondirectscalars}

The scattering amplitudes of the processes $e^{-}e^{+}\rightarrow
\phi^+ \phi^-$ and $e^{-}e^{+}\rightarrow \phi^{++} \phi^{--}$  are
dependent on the free parameters of littlest Higgs model $f,s,s'$.
In this work, the dependence of the cross sections on
mixing angles $s,s'$ at $f=1TeV$ is examined. For larger values of $f$, the
processes are not accessible for $\sqrt{s}\leq 3TeV$ because of the
kinematical limits of high masses of scalars. The processes are also
strongly dependent on mass of the Higgs scalar. In this work the input parameters are taken as: the Higgs mass $M_H=120GeV$ and the mass of the standard model gauge
bosons $M_{Z_L}=91GeV$, $M_{W_L}=80GeV$ and $s_W=0.47$ consistent
with recent data\cite{pdg}.
%%%%%%%%%%%%%%%%%%%%%%%%%%%%%

\begin{figure}[htb]
\begin{center}
\includegraphics[width=7cm]{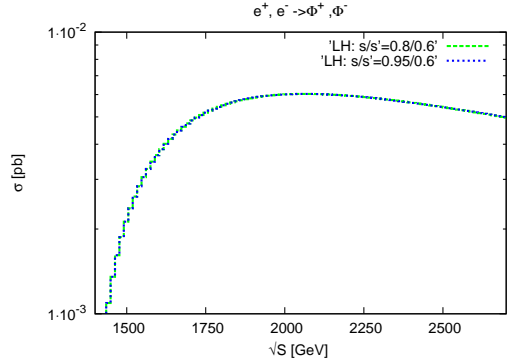}%\boss\includegraphics[width=5.5cm]{P1NSSp.eps}

 \caption{Total
cross section vs. $\sqrt{S}$ graphs for $e^{-}e^{+}\rightarrow
\phi^+ \phi^-$, for parameters  $s/s'$ when $f=1000GeV$.} \label{HPc1}
\end{center}
\end{figure}

The total cross sections for the single charged pair productions are
plotted in figure \ref{HPc1} with respect to center of mass
energy. Within the range of electroweak precision data, the direct
production cross section gets a value of $8\times10^{-3}pb$ for
$s/s'=0.8/0.6,0.95/0.6$ and $f=1TeV$ for energies $\sqrt{S}>1.7TeV$.
This will give up to $800$ productions per year for luminosities of
$100fb^{-1}$. For higher values of symmetry breaking scale parameter
($f>1TeV$), the production process is not accessible for ILC and
CLIC because of the kinematical constraints of high scalar mass.

At the final state, the decays of the single charged scalar pair
strongly depends on the values of the Yukawa couplings $Y$ and $Y'$
as seen from equation \ref{dwp1}. For $Y\simeq 1$ $\phi^\pm$
dominantly decays into same family leptons $l_i \nu_i (\bar{l_i}
\bar{\nu_i})$, violating lepton number. For small values $Y\ll 0.1$
the dominant decay mode is to SM pairs $W^{\pm}_L H$ and $W^{\pm}_L
Z_L$. The decay into different families of leptons $l_i \nu_j
(\bar{l_i} \bar{\nu_j})$ is proportional to $Y'^2$ and since the
flavor mixing Yukawa coupling $Y'$ is constrained to be small
no such signals are expected.

The final states of the single charged pair and their
production rates are calculated as:
\begin{itemize}
    \item $l_i \nu_i \bar{l_j}\bar{\nu_j}$ ~~~~:~~~~$PR(Y\simeq
    1)=\frac{2}{3}BR[Y]^2 \simeq 0.6 $,
    \item $l_i \nu_i W^{+}_L H (W^{+}_L Z_L)$ ~~~~:~~~~$PR(Y\simeq 0.2)=BR[Y] BR[SM] \simeq
    0.2$,
    \item to couples of SM, $W^{\pm}_L H$ and $W^{\pm}_L Z_L$ ~~:~~$PR(Y\ll 0.1)\simeq
    1$,
\end{itemize}
where $BR[SM]$ is the branching ratio to SM particles.

Finally, if $Y\simeq1$, about $500$ events per year can be observed at an
$e^+e^-$ collider with luminosity of $100fb^{-1}$ for $\sqrt{S}>1.7
TeV$. The final states will be $l_i \nu_i \bar{l_i}
\bar{\nu_i}$, which will be detected as two leptons and missing energy. Unfortunately this signal does not 
allow observing distinct features of the littlest Higgs model and lepton flavor violation, since the 
neutrino flavors can not be identified. Thus this channel contain high 
SM background, and these signals can have significance only if a detailed background analysis is done. 
%
%%%%%%%%%%%%%%%%%%%%%%%%

\begin{figure}[b]
\begin{center}
\includegraphics[width=7cm]{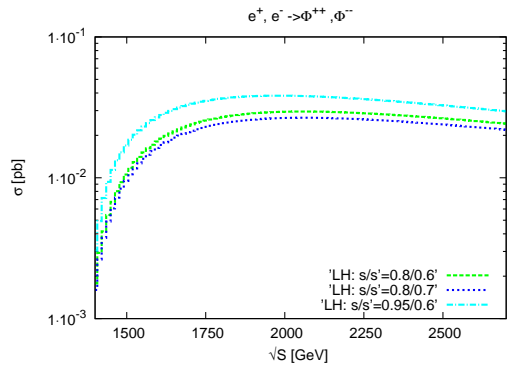}%\boss\includegraphics[width=5.5cm]{P2NSSp.eps}
%\vskip2mm (a)\hskip50mm(b) 
\caption{Total cross section vs.
$\sqrt{S}$ graphs for the process $e^{+}e^{-}\rightarrow \phi^{++} \phi^{--}$ for some values of 
mixing angle parameters $s/s'$ when $f=1000GeV$.} \label{HPc2}
\end{center}
\end{figure}

For the process $e^{+}e^{-}\rightarrow \phi^{++} \phi^{--}$ the total cross section of the production event is examined 
for $f=1TeV$. The dependence of the total cross section of the
$e^{+}e^{-}\rightarrow \phi^{++} \phi^{--}$ process on $\sqrt{s}$ at
fixed values of the littlest Higgs model parameters are presented in
Fig. \ref{HPc2}. For the mixing angles $s/s'=0.8/0.6$, the production cross section is maximum 
with a value of $3\times10^{-2}pb$ for $f=1TeV$ at $\sqrt{S}>1.7TeV$. 
For the parameter sets $s/s':0.8/0.7,0.95/0.6$ for $f=1TeV$, the total cross section is is slightly lower but still 
in the order of 
$10^{-2}pb$ at $\sqrt{S}>1.7TeV$. Thus for an $e^+e^-$ collider with an integrated luminosity of $100fb^{-1}$, yearly $3000$ double charged pair production can be observed. 

\begin{figure}[bh]
\begin{center}
\includegraphics[width=8cm]{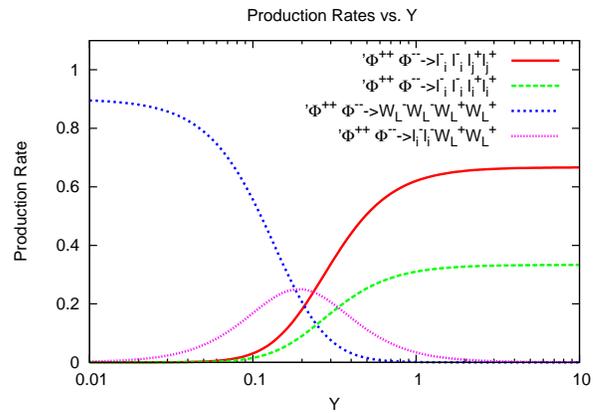}
 \vskip2mm \caption{Dependence of the final collider signatures of the process $e^+e^- \to \phi^{++}\phi^{--}$ on $Y$.} \label{BR2}
\end{center}
\end{figure}

It is seen from the final decay modes of the doubly charged pair given in Eq.\ref{dwp2} that the final state 
analysis is strongly dependent on the value of the Yukawa coupling $Y$, thus on the value of the triplet 
VEV $v'$. The dependence of the final collider signatures of the process $e^+e^- \to \phi^{++}\phi^{--}$ on $Y$ 
are plotted in figure \ref{BR2}.

For $Y<0.01(v'>10^{-8}GeV)$, the final decays of the doubly charged scalars are dominated by SM charged bosons, 
and the final collider signature will be $W^+_L W^+_L W^-_L W^-_L$. In this case with a subtraction from background,
 doubly charged scalars can be identified by reconstructing the same sign boson pairs.

For $0.01<Y<1(10^{-10}GeV<v'<10^{-8}GeV)$, the semi leptonic decay modes, $l_i l_i W_L^+ W_L^+$ will be observed. 
For $Y\sim 0.2$, the production rates for these modes are calculated as; $PR(Y\sim0.2)=0.2$, leading to $600$ collider signals 
per year at collider with a luminosity of $100fb^{-1}$ at $\sqrt{S}>1.7TeV$. The cleanest signal in this case 
will be observed when both $W_L^+$ decay into jets ($48\%$). In this scenario there will be yearly $280$ signal of
two same sign leptons of same family plus jets, violating lepton number and flavor by two, which can be directly detectable free from any backgrounds.

The most interesting scenario happens when the Yukawa coupling is close to unity, $Y\sim 1(v'\sim 10^{-10}GeV)$. In this case all final doubly charged 
scalars will decay into same family leptons, and the final collider signal will be either $l_i l_i \bar{l_j} \bar{l_j}$  or 
$l_i l_i \bar{l_i} \bar{l_i}$. The mixed states $l_i l_j \bar{l_i} \bar{l_j}$ resulting from the final decays of 
the doubly charged scalars into different families of leptons are suppressed because of the low production rates due to the value of the Yukawa mixing coupling $Y'$.

The branching ratio into the final signal $l_i l_i \bar{l_j} \bar{l_j}$ when $Y\sim1$ is calculated as $PR(Y\sim1)=0.66$, which will 
give $1800$ observable signals per year at an integrated luminosity of $100fb^{-1}$ when $\sqrt{S}>1.7TeV$. 
For this final signal, the lepton flavor is violated explicitly, free from any backgrounds.  The double charged scalars in this case 
can be reconstructed from invariant mass distributions of same charged leptons.

For the final state $l_i l_i \bar{l_i} \bar{l_i}$, there will be additional observable $1000$ events per year. For this case lepton flavor violation can not be observed 
directly even if it happens via the lepton flavor violating decays of doubly charged scalars. Although this signal has a huge SM background, with a proper background analysis, 
the existence of doubly charged scalars and so on their lepton flavor violating decays can be identified.

%\clearpage

In conclusion, it is found that pair productions of single and doubly charged scalars are in the reach for $e^+ e^-$ colliders when $\sqrt{S}\geq 1.7TeV$. The production rates of 
the single charged pair are quite low and final signatures are not promising since neutrinos in the final states 
can not be identified. For the pair production of doubly charged scalar via process  $e^{-}e^{+}\rightarrow \phi^{++} \phi^{--}$, the production 
rates are promising and also the final state signatures are  distinct observable collider signals depending on final decay modes of the doubly charged pair which were discussed in\cite{thanlept1}. 
When $0.01<Y<1$, the final state will be semileptonic; i.e: jets plus two same sign leptons of 
same family, giving an explicit signature of lepton number violation by two. And also 
for the case when $Y\sim 1$, the final decays of the doubly charged pair will be leptonic, 
including an explicit lepton flavor violation in the final state $l_i l_i \bar{l_j} \bar{l_j}$. These both 
final signatures are due to distinct features of littlest Higgs model and 
will be free from any backgrounds.

\clearpage
\begin{acknowledgments}
A. \c{C} thanks to T.M Aliev and A. \"Ozpineci for their guidance and help
during this work.
\end{acknowledgments}
%
%\newpage
%
%%%%%%%%%%%%%%%%%%%%%%%%%%%%%%%%%%%%%%%%%%%%%%%%%%%%%%%%%%%%%%%%%%%%%%%%%%%

\clearpage
\end{document}